\documentclass[sn-mathphys-ay]{sn-jnl}

\usepackage{lmodern} 
\usepackage{graphicx}%
\usepackage{multirow}%
\usepackage{amsmath,amssymb,amsfonts}%
\usepackage{amsthm}%
\usepackage{mathrsfs}%
\usepackage[title]{appendix}%
\usepackage{xcolor}%
\usepackage{textcomp}%
\usepackage{manyfoot}%
\usepackage{booktabs}%
\usepackage{algorithm}%
\usepackage{algorithmicx}%
\usepackage{algpseudocode}%
\usepackage{listings}%

\begin{document}

\title{The Trade-Off between Directness and Coverage\\ in Transport Network Growth}

\author*[1]{\fnm{Clément} \sur{Sebastiao}}\email{clse@itu.dk}

\author[1,2]{\fnm{Anastassia} \sur{Vybornova}}

\author[3]{\fnm{Ane Rahbek} \sur{Vierø}}

\author[1,4]{\fnm{Luca Maria} \sur{Aiello}}

\author[1,4,5]{\fnm{Michael} \sur{Szell}}

\affil[1]{\orgdiv{NEtwoRks, Data, and Society (NERDS)}, \orgname{IT University of Copenhagen}, \orgaddress{\street{Rued Langgaards Vej 7}, \city{Copenhagen}, \postcode{2300}, \country{Denmark}}}

\affil[2]{\orgdiv{Copenhagen Center for Social Data Science (SODAS)}, \orgname{Copenhagen University}, \orgaddress{\street{{\O}ster Farimagsgade 5}, \city{Copenhagen}, \postcode{1353}, \country{Denmark}}}

\affil[3]{\orgdiv{Mobility, Space, Place and Urban Studies (MOSPUS)}, \orgname{Roskilde University}, \orgaddress{\street{Universitetsvej 1}, \city{Roskilde}, \postcode{4000}, \country{Denmark}}}

\affil[4]{\orgname{Pioneer Centre for AI}, \orgaddress{\street{{\O}ster Voldgade 3}, \city{Copenhagen}, \postcode{1350}, \country{Denmark}}}

\affil[5]{\orgname{Complexity Science Hub}, \orgaddress{\street{Metternichgasse 8}, \city{Vienna}, \postcode{1030}, \country{Austria}}}

\abstract{Designing spatial networks, such as transport networks, commonly deals with the problem of how to best connect a set of locations through a set of links. In practice, it can be crucial to order the implementation of the links in a way that facilitates early functioning of the network during growth, like in bicycle networks. However, it is unclear how this early functional structure can be achieved by different growth processes. Here, we systematically study the growth of connected planar networks, quantifying functionality of the growing network structure. We compare random growth with various greedy and human-designed, manual growth strategies. We evaluate our results via the fundamental performance metrics of directness and coverage, finding non-trivial trade-offs between them. Manual strategies fare better than greedy strategies on both metrics, while random strategies perform worst and are unlikely to be Pareto efficient. Centrality-based greedy strategies tend to perform best for directness but are worse than random strategies for coverage, while coverage-based greedy strategies can achieve maximum global coverage as fast as possible but perform as poorly for directness as random strategies. Directness-based greedy strategies get stuck in local optimum traps. These results hold for a number of stylized urban transport network topologies. Our insights are crucial for applications where the order in which links are added to a spatial network is important, such as in urban or regional transport network design problems.}


\maketitle

\section{Introduction}
Transport network design is a complex endeavor that can consist of a plethora of different tasks and considerations \citep{farahani2013review}, benefiting from different perspectives such as quantitative geography, transport engineering, or physics \citep{xie_modeling_2009}. A planar spatial network already underlies strong spatial limitations \citep{barthelemy_spatial_2022}, to which a transport function adds additional constraints such as providing direct connections, serving a given population, avoiding bottlenecks \citep{kirkley2018betweenness}, or multimodality \citep{alessandretti2023multimodal}. 

Designing networks by optimization with constraints naturally comes with trade-offs. Such trade-offs have been explored in general network growth processes, such as average shortest path versus link density \citep{cancho2003optimization} or total length \citep{brede2010coordinated}, costs versus benefits \citep{louf_emergence_2013}, or comparing measures of efficiency \citep{latora2002boston,cardillo_structural_2006,morer_comparing_2020}. More particularly in transport, different applications come with different trade-offs, such as directness versus coverage in bicycle networks \citep{szellGrowingUrbanBicycle2022}, densification versus exploration in road networks \citep{strano_elementary_2012, el_gouj_urban_2022}, centralization in air transport networks \citep{wangEvolutionAirTransport2014}, or growing versus shrinking in rail networks \citep{bottinelli_efficiency_2019}. 

A specific formulation of a transport network design problem is underexplored so far, emerging from recent studies of bicycle network growth \citep{szellGrowingUrbanBicycle2022,steinacker_demand-driven_2022,vybornova_automated_2022,paulsen_societally_2023,mauttone_bicycle_2017,ospina_maximal_2022,steinacker2025robust}. The starting point in this context is either to consider all links of the street network for potential future protected bicycle infrastructure, or a subset of all links that has been identified by planners to connect a number of locations. In short, the plan for a future, final network exists. However, once identified, the network does not appear immediately but needs to be constructed gradually. As construction can take decades, it is crucial to implement the connections in an order that serves cyclists as early as possible. Although cyclists can legally use the underlying street network, they are typically not able to do so safely, so a delay in building protected bicycle infrastructure should be avoided as it could severely obstruct the uptake of cycling \citep{schoner2014missing,ferenchak2025link}. The links of protected bicycle infrastructure that will be implemented along the road network therefore need to be functional to cyclists as soon as possible, for example to provide uninterrupted or direct routes. While this problem is motivated by the bicycle network context, it is a special case of a research question that can be formulated more generally, where ``functionality'' depends on the particular use case:

\begin{quote}
Given a set of locations (nodes) and a set of possible connections (links) between them, in which order should these connections be added to ensure a functional network structure already during growth? 
\end{quote}
This research question is especially important for sustainable infrastructure networks like bicycle networks, since high political friction and extremely limited funding due to car-centrism imply long construction times \citep{carstensen_spatio-temporal_2015} that can leave network gaps open for many decades \citep{nateraorozcoDatadrivenStrategiesOptimal2020, vybornova_automated_2022}. Together with the urgency of the climate crisis, there is thus a pressing need to understand those designs that arrive at a \emph{functional} network as soon as possible, i.e., \emph{a network that is useful for its users throughout the construction process -- the earlier the better}. As the network should be used already during its growth stages, in this paper we focus not on finding the optimal final shape of the network, but we assume the final possible shape as given and consider the optimality of use \emph{during growth}. Further, we operationalize ``functionality'' via the two performance metrics of directness and coverage, see Section \ref{method}.

The posed network design question has been addressed from both transport planning using multi-parameter optimization \citep{mauttone_bicycle_2017,ospina_maximal_2022,paulsen_societally_2023,ballo2024designing,wiedemannBikeNetworkPlanning2025} and from network science using single or multi-parameter optimization \citep{steinacker_demand-driven_2022,szellGrowingUrbanBicycle2022,steinacker2025robust}. In several such previous studies the growth algorithm is greedy, adding only the best possible next link without considering the long-term future. Further, instead of such an additive approach, \cite{steinacker_demand-driven_2022,ballo2024designing,steinacker2025robust,wiedemannBikeNetworkPlanning2025} have explored a \emph{subtractive} design process where starting backwards from the final network the least important link is removed in each step. While all these approaches provide useful applications, so far a \emph{systematic} study of spatial network growth processes and the evolution of their performance metrics is missing. Such a study is the contribution of our paper. In particular, we study the growth of connected planar networks and explore the trade-off between the two most important metrics, directness and coverage.

The rest of the paper is structured as follows. Section \ref{method} introduces the methodology, describing the network metrics used as cost functions during the growth, and the overall framework for optimal growth of planar networks. Section \ref{results} presents the results of applying these methods to one of the simplest planar networks -- a regular grid lattice -- and then extends the analysis to other stylized urban street patterns. Section \ref{sec:discussion} discusses the results and puts them into context. The method developed here is available as reproducible open-source code at \cite{github_ob} which can be applied to any network structure or metric.

\section{Methods}
\label{method}

\subsection{Strategies and orders}

In this article we study the concept of link ordering \emph{strategy}, which we define as an algorithm $S$ that takes as input the spatial network $\mathcal{G}=(\mathcal{N},\mathcal{L})$ made up of the set of nodes $\mathcal{N} = \{n_1,\ldots,n_N\}$, and links $\mathcal{L} = \{e_1,\ldots,e_L\}$. We call this network $\mathcal{G}$ the \emph{final network} -- it can represent a future network that planners are deliberately targeting to build, or a network with a large number of links which could be implemented potentially. The strategy outputs a sequence (ordered list) of links encoding in which order they should be implemented: $S(\mathcal{G}) = (\sigma(e_i))_{i=1}^{L}$, where $\sigma$ is a permutation determined by $S$. We call $(\sigma(e_i))_{i=1}^{L}$ a (link) \emph{order} of strategy $S$ for the network $\mathcal{G}$. For example, a strategy might determine that the best order for a final network $\mathcal{G'} = (\{n_1,n_2,n_3\}, \{e_1,e_2,e_3\})$ is the sequence $(e_3,e_1,e_2)$, in which case link $e_3$ should be implemented first, followed by link $e_1$, followed by link $e_2$. We call a subnetwork $\mathcal{H}=(\mathcal{N},S(\mathcal{G})_{i=1}^{k}) \subseteq \mathcal{G}$ that has built the first $k$ links according to strategy $S$ the \emph{$k$th step} of network $\mathcal{G}$ for strategy $S$.

If a strategy's output is always the same for the same input, it is fully deterministic. However, in our study, each considered strategy has multiple possible orders for the same set of nodes and links because randomness is always part of the process -- either by construction in a random strategy, or when a new added link has to be selected randomly among multiple equivalent link choices. So, for the example strategy above, calling the algorithm another time on the same final network could yield a different order $(e_1,e_3,e_2)$. To observe the variation of strategies, we create 50 orders for each strategy, except for the random strategy where we create 1000 orders as it generates wider variations.

\subsection{Network metrics}
\label{netmet}

We use several network metrics for two purposes: 1) to evaluate the quality of a given link order, and 2) for defining greedy growth strategies. Our framework works with any metric, but here we pick specific ones that are most relevant for the bicycle network link ordering problem motivated in the introduction. The formalism is general though and could be applied to a wide range of other use cases, such as designing road, rail or other spatial networks. Further, the networks we consider are planar by construction because urban street networks are mostly planar \citep{boeing2020planarity}, but strict planarity is not a necessary prerequisite. It is important though that the networks are spatial.

The Dutch CROW Manual \citep{CROW_manual}, a key reference for bicycle infrastructure design, reports four main requirements for a good bicycle network: cohesion, directness, safety, and attractiveness. While safety and attractiveness are mostly about the quality of the physical infrastructure and its surroundings, cohesion and directness relate to the topology of the bicycle network. Cohesion was operationalized by \cite{szellGrowingUrbanBicycle2022} as being able to reach all points of interest (accessibility, or coverage) in an interconnected manner (connectedness). Directness is the ratio between shortest path distances and beeline distances.

\begin{figure}
    \includegraphics[width=\linewidth]{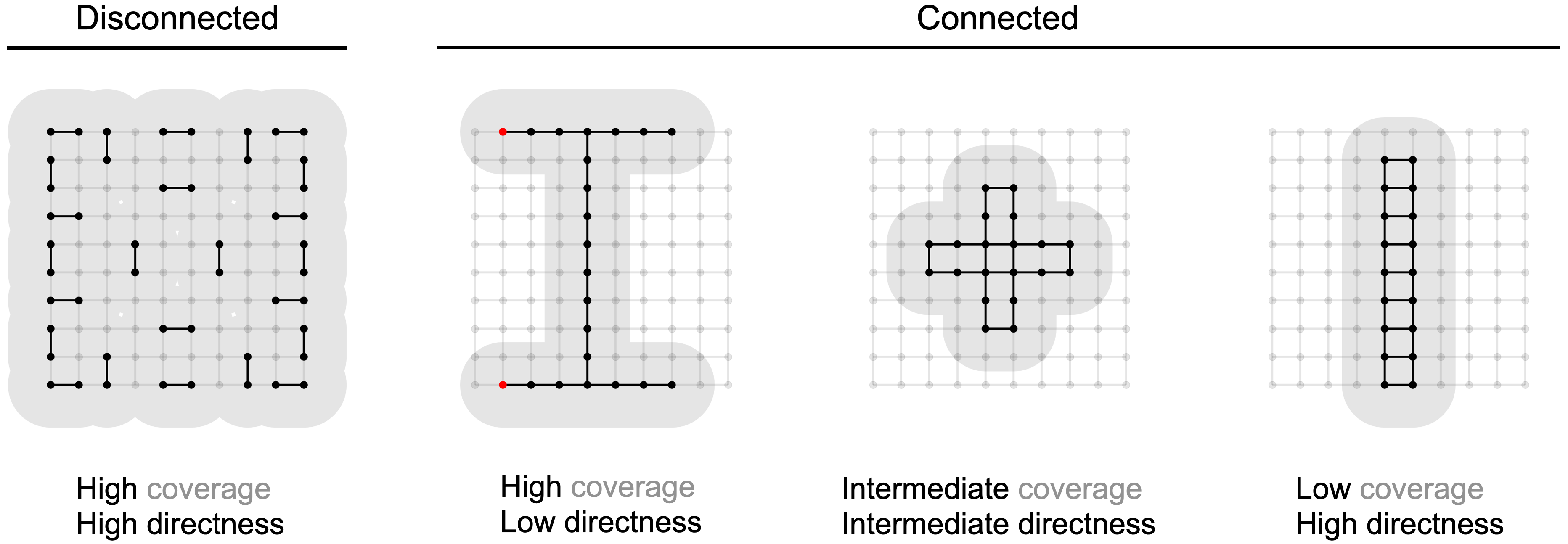}\\
    \caption{\textbf{The directness-coverage trade-off.} Four examples of growing networks at step 24 are illustrated for the final network of the $10\times10$ grid (gray network). Gray areas denote buffered network elements which determine coverage. From left to right: Growth snapshot from 1) a disconnected strategy with high coverage and directness. Due to being disconnected, this strategy is not useful for most applications; 2) a connected strategy with high coverage and low directness -- the red nodes highlight one of the large detours when directness is not prioritized; 3) a connected strategy with intermediate coverage and directness, exemplifying a trade-off between the two metrics; 4) a connected strategy with low coverage and high directness.}
    \label{fig:explainer}
\end{figure}

As we do not deal with spatial attributes other than distances between nodes, we are dropping the safety and attractiveness aspects in the following. We focus on the topological/geometrical aspects of coverage, directness, and connectedness for measuring performance of our growth process, defined in detail in the following subsections. Figure~\ref{fig:explainer} illustrates the specific multi-criteria optimization that we address. We aim to design a growing network where both coverage and directness are high, but we also expect a trade-off between the two. For example, previous work has shown that growing disconnected networks via a random strategy leads to high coverage at the expense of other network metrics \citep{szellGrowingUrbanBicycle2022}. 

At the same time, we impose the condition of connectedness through all growth stages (i.e., to consist of one single connected component), because disconnected components are an undesirable trait, especially in transport networks \citep{nateraorozcoDatadrivenStrategiesOptimal2020}. 
Further, greedy network growth would not be meaningful when isolated links or components were allowed, as it would be trivial and not desirable, see leftmost panel of Fig.~\ref{fig:explainer}: A growth process in which isolated pairs of nodes connected by single, straight links are continuously added to the network would optimize both directness and coverage. In this case, the optimization of directness happens because the only existing origin-destination pairs are neighboring nodes; the optimization of coverage happens because the buffers around scattered links do not overlap, thus contributing their full area while the network is sparse. The connectedness condition implies that we will never create a new connected component when growing the network. The fundamental performance metric of connectedness is therefore fulfilled by construction and will not be considered further.

Besides the two relevant performance metrics of coverage and directness, we consider several other optimization metrics in a greedy process where we grow by optimizing the optimization metric step by step, i.e., link by link. 
As we explore the growth of \emph{connected} planar networks from scratch, potential performance metrics for disconnected networks such as the number of connected components, the length of the largest connected component, or the relative directness, are not relevant \citep{szellGrowingUrbanBicycle2022}. The two selected performance metrics of coverage and directness coincide with two of our optimization metrics: For example, we can grow by optimizing directness step by step and observe the effect on directness itself, expecting high performance that is locally optimal by construction but may or may not be globally optimal. In addition, we consider optimization metrics that are often used for spatial networks, like closeness and betweenness centrality \citep{barthelemy_spatial_2022}. Some of these metrics allow a direct comparison with previously studied growth strategies of planar networks \citep{szellGrowingUrbanBicycle2022}. We define all the performance and optimization metrics in the following subsections.

\subsubsection*{Coverage}

Coverage is the area covered by the spatially buffered links of the network, see gray areas in Fig.~\ref{fig:explainer}. For buffering we use rounded line endings, which gives 
\begin{align}
A(l,d) &= 2ld + 2\frac{1}{2}\pi d^2\notag\\
    &= \left(\frac{2l}{d} + \pi\right)d^2\label{eq:bufferedlink}
\end{align}
for the area of a buffered link of length $l$ with buffer size $d$. Buffer size heavily influences the results and must be chosen reasonably: For a buffer size too small, $d\ll l$, all links account for roughly the same amount of added coverage; for a buffer size too large, $d\gg l$, there is too much buffer overlap and links that do not lie on the periphery of the network will not account for coverage. In both these cases, little to no information would be gained, which makes the choice of an intermediate buffer size crucial.

If coverage in this simple, purely areal definition was interpreted as a proxy for coverage of certain spatial elements like a population or locations of interest, it would assume a homogeneous distribution of such elements. Thus, additional refinements could make the coverage concept more realistic in specific applications, such as accounting for an empirical, non-homogeneous distribution of the spatial elements that are aimed to be covered. 

As our chosen lengths $l$ are arbitrary, in the following we report coverage in a normalized form where a normalized coverage value of 1 means that the final network's covered area is reached.

\subsubsection*{Adaptive coverage}

As explained above, the buffer size should be intermediate. However, choosing a single constant buffer size cannot account for different network scales. For example, road networks are typically hierarchical, having different meaningful scales \citep{batty2006hierarchy}, which also holds for other networks such as the cardiovascular system or the branching structure of a tree \citep{west2005origin}. 

This issue can be addressed with adaptive coverage, where the buffer changes throughout the growth process. Starting with a large buffer size that is decreased over time, different scales can be covered iteratively. In the context of human transport networks, this idea allows building first a ``network of main roads'', i.e. a sparse network covering the whole area crudely, and then iteratively adding refined ``side roads'' that cover more and more nooks.

To implement adaptive coverage, we start from a large buffer size and reduce it whenever an added link contributes less than a given threshold to the area of the buffered network. As mentioned previously, the area of a buffered link of length $l$ with a buffer size $d$ is $A(l,d) = (\frac{2l}{d} + \pi)d^2$. As our growth process imposes connectedness, with each new link there will be necessarily a minimum share of the new link's buffered area with the existing coverage, which is the area of the circle $\pi d^2$ around at least one existing connecting node. We therefore consider the ratio $r = \frac{\Delta A}{A(l,d)-\pi d^2}$ between the area $\Delta A$ that is effectively added and the area ${A(l,d)-\pi d^2}$ of the buffered link minus the necessary overlap $\pi d^2$. If at any step the ratio $r$ falls below a specified threshold (which we set to 0.1), the buffer size is halved for the next step. This procedure is adequate if the network intuitively resembles a simplified road network, i.e., if nodes are not too clustered but roughly homogeneously distributed in space. For this reason, in practical applications to transport networks, geometric or topological preprocessing would be required to simplify areas with high auxiliary node densities such as roundabouts or large intersections \citep{boeing2017osmnx,boeing2025topological,fleischmann2025adaptive}.

\subsubsection*{Directness}

We compute directness $D$ as the average of the ratio between Euclidean distance $d_E(i,j)$ and shortest network path length $d_G(i,j)$ over all node pairs $i$ and $j$:

\begin{equation}
    D =  \left< \frac{d_E(i,j)}{d_G(i,j)} \right>_{i \neq j}
\end{equation}

\noindent If $D=1$, any path between any pair of nodes is straight, i.e. identical with the beeline. In other words, the more direct a network is, the shorter the path from any node to any other node. In the literature, this value (or the inverse of this value) can be found under many other names: detour factor, detour index, detour ratio, circuity, route factor, and straightness \citep{CROW_manual,reggianiUnderstandingBikeabilityMethodology2021,barthelemy_spatial_2022}. From a geometric perspective, $D$ is also equivalent to the average sinuosity of the shortest paths. Directness is typically used to determine how efficient a spatial network is, as the closer $D$ is to 1, \emph{ceteris paribus}, the faster travelers will reach their destinations. The choice of the origins and destinations impacts the final estimation. Some studies take random origins to sample the directness \citep{giacominRoadNetworkCircuity2015}, others use trajectories from real trips to determine the typical directness experienced \citep{yangUniversalDistributionLaw2018}. For computations of accessibility, another way of finding meaningful origins and destinations is selecting households and amenities, respectively \citep{lowryPrioritizingNewBicycle2016}. Since here we work with synthetic networks, and do not want to depend on an arbitrary choice of selected origin-destination pairs, we consider the all-to-all network between all pairs of nodes in line with the literature \citep{vybornova_automated_2022}. For a grid of size $N \times N$, we thus compute directness $D$ as:

\begin{equation}
    D = \frac{\sum\limits_{i=1}^{N}\sum\limits_{j=1}^{N}\sum\limits_{k=1}^{N}\sum\limits_{l=1}^{N}d(i, j, k, l)}{N^2(N^2-1)}
\end{equation}

\noindent where

\begin{equation}
    d(i, j, k, l) = 
\begin{cases}
    0 & \text{if } (i,j)=(k,l)\\
    \frac{\sqrt{|k-i|^2+|l-j|^2}}{|k-i|+|l-j|} & \text{otherwise}
\end{cases}
\end{equation}
The CROW Manual \citep{CROW_manual} recommends a bicycle network to have a directness value of at least $D = 1/1.26 \approx 0.79$. For a grid, the directness decreases and converges with increasing size to $D\approx0.79$, which coincides with the recommended minimum value.

\begin{figure}
    \centering
    \includegraphics[width=0.8\textwidth]{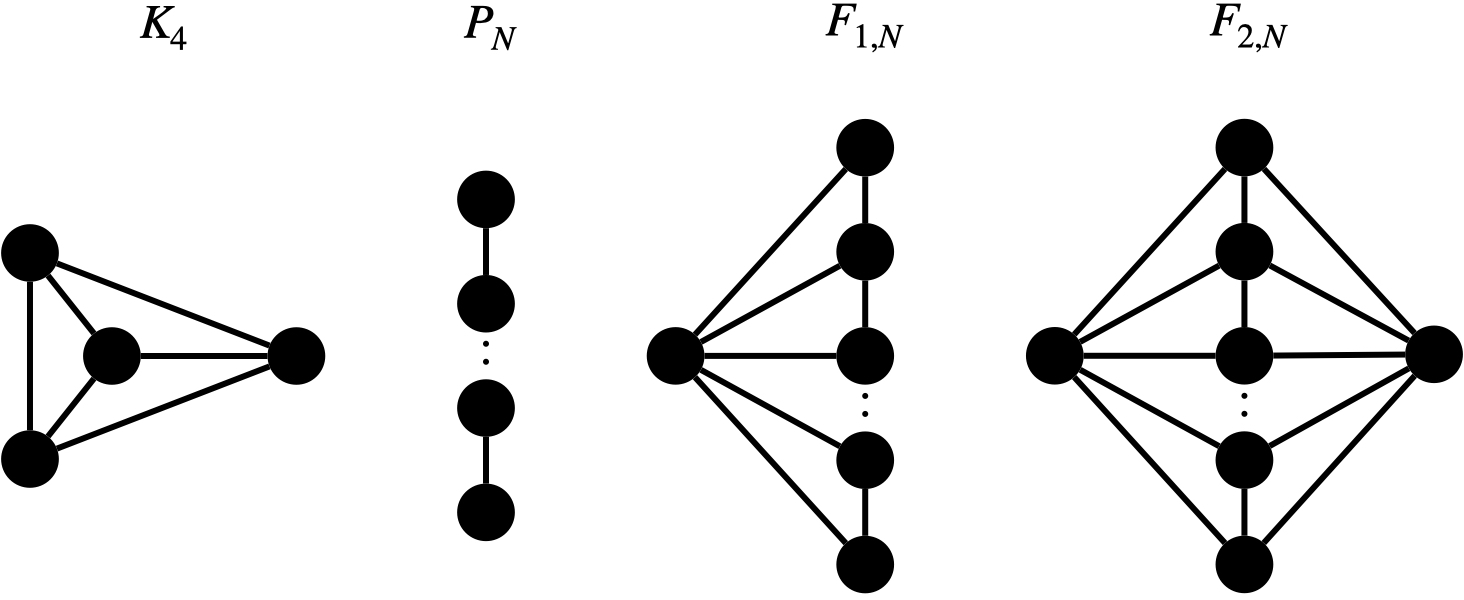}
    \caption{\textbf{The set of all planar networks with directness $D=1$.} $K_4$: The fully connected network with four nodes; $P_N$: The family of path networks  of arbitrary length $N$, which includes $K_2$; $F_{1, N}$: The family of one-sided fan networks, which includes $K_3$; and $F_{2, N}$: Two-sided fan networks.}
    \label{max_dir_graph}
\end{figure}

Directness is highly non-linear as it depends on the relative position of each node to all other nodes. The only ways in which a network can have the maximum directness $D=1$ is either by being fully connected, by having only straight paths, or by a limited combination of these properties. For planar networks, which transport networks are to a large extent \citep{boeing2020planarity}, these are, Fig.~\ref{max_dir_graph}, the fully connected network with four nodes $K_4$; the family of path networks $P_N$ of arbitrary length $N$, which includes $K_2$; and the family of one/two-sided fan networks $F_{1/2, N}$ which are the unions of one/two nodes and a path network $P_N$ \citep{brandstadt1999graph}. One-sided fan networks $F_{1, N}$ include $K_3$. All other $K_N$ for $N>4$ are not planar \citep{kuratowski1930probleme}. Additional non-intersecting, curved links could be added to $P_N$ or to $F_{1/2, N}$ until reaching maximal planarity, but would be irrelevant as $D$ only considers lengths of shortest paths.

While coverage is monotonously non-decreasing with growth, this behavior is different for directness. A larger network will usually have a lower directness than a smaller one, as it is harder to keep a planar network close to being fully connected (and thus direct) than a smaller one: a complete network $K_N$ has $\frac{N(N-1)}{2} \sim N^2$ links, while a maximally planar network has only $3N-6 \sim N$ links.

\subsubsection*{Relative directness}

We want to evaluate the different orders in which a fixed set of links can be added to a fixed set of nodes. Thus, the directness for each node pair is capped to the directness for the node pair in the final network. From this observation, we compute the relative directness $D_R$, where instead of comparing the Euclidean distance to the shortest network path length, it compares the actual shortest network path length $d^A_G(i,j)$ to the final shortest network path length $d^F_G(i,j)$:

\begin{equation}
    D_R =  \left< \frac{d^A_G(i,j)}{d^F_G(i,j)} \right>_{i \neq j}
\end{equation}
If $D_R=1$, the path between any randomly selected pair of nodes is as short as it is on the final network.

Out of all considered metrics in this study, $D$ and $D_R$ are the most computationally demanding ones, because the shortest path computations, already computationally expensive in themselves, have to be conducted for every connected pair of nodes, for every tested link, at every growth step, yielding a computational complexity of $O(L^5)$ where $L$ is the number of links.

\subsubsection*{Closeness}
Closeness centrality \citep{sabidussiCentralityIndexGraph1966} measures how close (in terms of shortest path length) a node $u$ is to any other node of the network,
\begin{equation}
    C_C(u) = \frac{N-1}{\sum\limits^{N-1}_{u \neq v}d(u,v)}
\end{equation}
where $N$ is the total number of nodes, and $d(u,v)$ is the shortest path length between nodes $u$ and $v$. Since the closeness centrality is only defined on nodes and we need to optimize by links, we compute the link closeness centrality $C^E_C(e)$ as the mean of the closeness of the nodes $(e_1, e_2)$ connected by the link $e$:

\begin{equation}
    C^E_C(e) = \frac{C_C(e_1) + C_C(e_2)}{2}
\end{equation}

\subsubsection*{Betweenness}
Link betweenness centrality \citep{girvanCommunityStructureSocial2002} measures how frequently a link $(i,j)$ is included in the shortest paths between any pair of nodes of the network,
\begin{equation}
    C^E_B(e) = \sum\limits_{s \neq t}\frac{\sigma_{st}(e)}{\sigma_{st}} \frac{2}{N(N-1)}
\end{equation}
where $\sigma_{st}$ is the number of shortest paths between node $s$ and node $t$, and $\sigma_{st}(e)$ is the number of shortest paths between node $s$ and node $t$ that include the link $e$. The normalization factor $\frac{2}{N(N-1)}$ is the number of node pairs $\binom{N}{2}$.

Assuming an equal number of travellers coming from and going to every node, the link betweenness centrality is a proxy for traffic flow and the reason why it is widely used in street network analysis \citep{barthelemy_spatial_2022}.

\subsection{Growth strategies}

\subsubsection*{Random growth}
\label{random-growth}
The simplest growth strategy is a random ordering of links. On the one hand, this strategy serves as a null model, or a lower bound, for other strategies. On the other hand, random growth can be a reasonable model for the observed growth of certain networks such as bicycle networks \citep{szellGrowingUrbanBicycle2022}.

For all random growth strategies we start growing at the link with highest closeness centrality, which, for comparability, is the same seed link for all growth strategies. Then, we consider adding a link randomly from all possible links from the final network that would preserve connectedness (due to the connectedness condition). We follow this procedure at each step until all links from the final network have been added.

\subsubsection*{Greedy optimization}
\label{greedy-optim}

\begin{algorithm}[t]
\caption{Compute the optimal order of growth of the network $G$ for the metric $f(G)$}
\begin{algorithmic}
    \Require{Planar network $G$, network metric $f(G)$}
    \State{Initialize $G_1$ with the link of highest closeness}
    \For{(\# of links in $G$) - 1}
    \State{Find the list of links $l \in G$ connected to $G_i$.}
    \For{\textbf{e} in $l$}
    \State{Add \textbf{e} to $G_i$, producing $G^\mathbf{e}_i$}
    \State{Compute $f(G^\mathbf{e}_i)$}
    \EndFor
    \State{Add \textbf{e} to $G_i$ for which $f(G^\mathbf{e}_i)$ is maximal}
    \EndFor
    \State \Return{Order of the links added}
\end{algorithmic}
\label{pseudocode}
\end{algorithm}

Greedy optimization considers at each step the optimal choice for that particular step only. It means always choosing the local optimum, not necessarily leading to the global optimum. See Algorithm~\ref{pseudocode} for how we implement our greedy optimization. When multiple links provide the same maximal value, we choose one among them randomly.

One point of our study is to show the pros and cons of greedy strategies relative to random and manual strategies, and whether it is possible to ``get away with'' a greedy strategy instead of trying to solve the network growth problem with a more advanced heuristic or a global approach \citep{wiedemannBikeNetworkPlanning2025}.

The advantages of greedy optimization are its ease of use and implementation, and its explainability. The latter is crucial for applications of the method, such as support in urban planning. The disadvantages of greedy optimization are well known \citep{cormen2022introduction}. As it is a simple method taking very limited knowledge into account, it can easily give suboptimal results due to falling into local optimum traps, especially for highly non-linear cost functions. 

We show here the results for what we call an ``additive'' strategy, also called ``forwards percolation'', ``forward'' or ``bottom-up'' strategy \citep{ballo2024designing,steinacker2025robust,wiedemannBikeNetworkPlanning2025}, where we start from an initial network and grow it by adding the link which yields the network with the maximal chosen optimization metric at each step until reaching the final network. Apart from that, we study a subtractive strategy, also called ``backwards percolation'', ``reverse'' or ``top-down'' strategy \citep{ballo2024designing,steinacker2025robust,wiedemannBikeNetworkPlanning2025}, where we instead start from the final network and remove the link which yields the network with the maximal chosen optimization metric at each step until reaching a single link. In Section \ref{results} we show results for the additive strategy as it usually outperforms the subtractive strategy (see Section \ref{add-vs-sub}).

\subsubsection*{Manual growth orders}

We want to compare our results from the greedy growth with other growth methods. Overall, these methods fall into three classes: random strategies as a lower bound; greedy strategies analyzed for each metric described in Section~\ref{netmet}; and manual orders, introduced below.

\begin{figure}
    \centering
    \includegraphics[width=\linewidth]{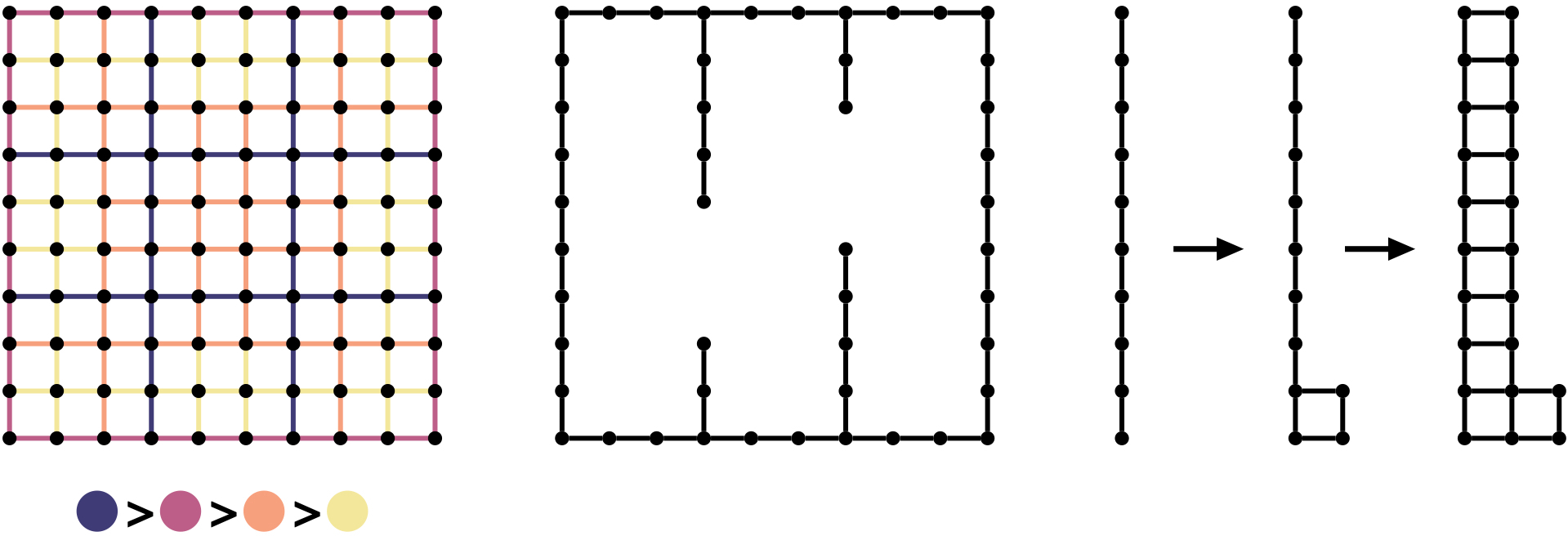}
    \caption{\textbf{The three manual growth orders:} Left: Manual order H following an implicit hierarchy of the network, from blue to pink to orange to yellow. Center: Manual order C optimizing for coverage. Shown here is step 48 out of 180, when reaching the full coverage. Right: Manual order D optimizing for directness.}
    \label{fig:manual}
\end{figure}

The general idea behind manual ordering is the incorporation of design strategies that imagine a human planner who aims to implement a global optimum for specific metrics, or certain macroscopic structures first such as a ``ring road'' or other hierarchical structures \citep{barthelemy_self-organization_2013,walker2024human}, being able to circumvent local optimum traps, see Fig.~\ref{fig:manual}. Thus, manual orders can serve as an upper bound for single metrics, or as an implementation of a globally optimal or ``human-designed'' strategy.

Since here we focus on directness and coverage, we manually design globally optimal growth orders for each of these two performance metrics, providing the best growth we can expect. This way we expect to discover how far away our greedy strategies are from the global optimum. We also design an intermediate ``hierarchical order'', explained below. We are able to design globally optimal manual orders for the regular grid, but not for other test networks, as their added complexity and inflated solution space make it infeasible to argue for global optimality. For comparability with the other growth strategies, in each of our manual orders we keep the constraint of connectedness, and we start with a link similarly central as for the random and greedy strategies -- we have verified that growth processes remain comparable.

\textbf{Manual hierarchical order H.}
\label{man-hier-ord}
We design a manual hierarchical order ``H'' to represent what a planner could consider a good hierarchical growth strategy. Without any additional knowledge of links, this order is a good compromise between different hard-to-define factors a planner could have in mind, i.e., balancing coverage and directness through iteratively refined macroscopic structures (like ``ring roads'', etc.). We create different levels of priority from the implicit hierarchy of the network (see left panel of Fig.~\ref{fig:manual}) and start by building from the highest hierarchy to the lowest, keeping the growth symmetrical. This growth order is based on human intuition and not on rigid optimization of metrics.

\textbf{Manual coverage order C.}
\label{man-cov-ord}
The manual coverage order ``C'' is providing the upper bound for coverage, for the fixed buffer size of 1.5 times the link length. Until reaching full coverage, this order follows closely greedy optimization for coverage. Once the full coverage is reached, as in the central panel of Fig.~\ref{fig:manual}, the growth continues following manual order H, by first connecting the two middle lines, then finishing the purple and magenta parts in the left panel of Fig.~\ref{fig:manual}, and then following the same strategy as ``H''.

\textbf{Manual directness order D.}
The manual directness order ``D'' is providing the upper bound for directness. We grow with the following pattern: we start by a straight line, as it is a network with maximum directness (see Fig.~\ref{max_dir_graph}). We then add the second line square by square, until reaching the end of the second line, and continue the same way until completing the grid (see right panel of Fig.~\ref{fig:manual}).

\section{Results}
\label{results}

We first describe our results on a square grid. Without loss of generality, we select a square grid with $10\times10$ nodes where each link is set to be 100 meters long, results are then reported in length units of kilometers. This choice is arbitrary but taken for the best correspondence with transport network applications. The chosen buffer size for both the Coverage strategy and for all subsequent analysis is 150 meters, or 1.5 times the length of a link.
For the Adaptive coverage strategy, we initiate with a buffer size of 300 meters, halving it to 150 and then again to 75 meters (see Section~\ref{var-can-met} for more details).

\begin{figure}
    \centering
    \includegraphics[width=\linewidth]{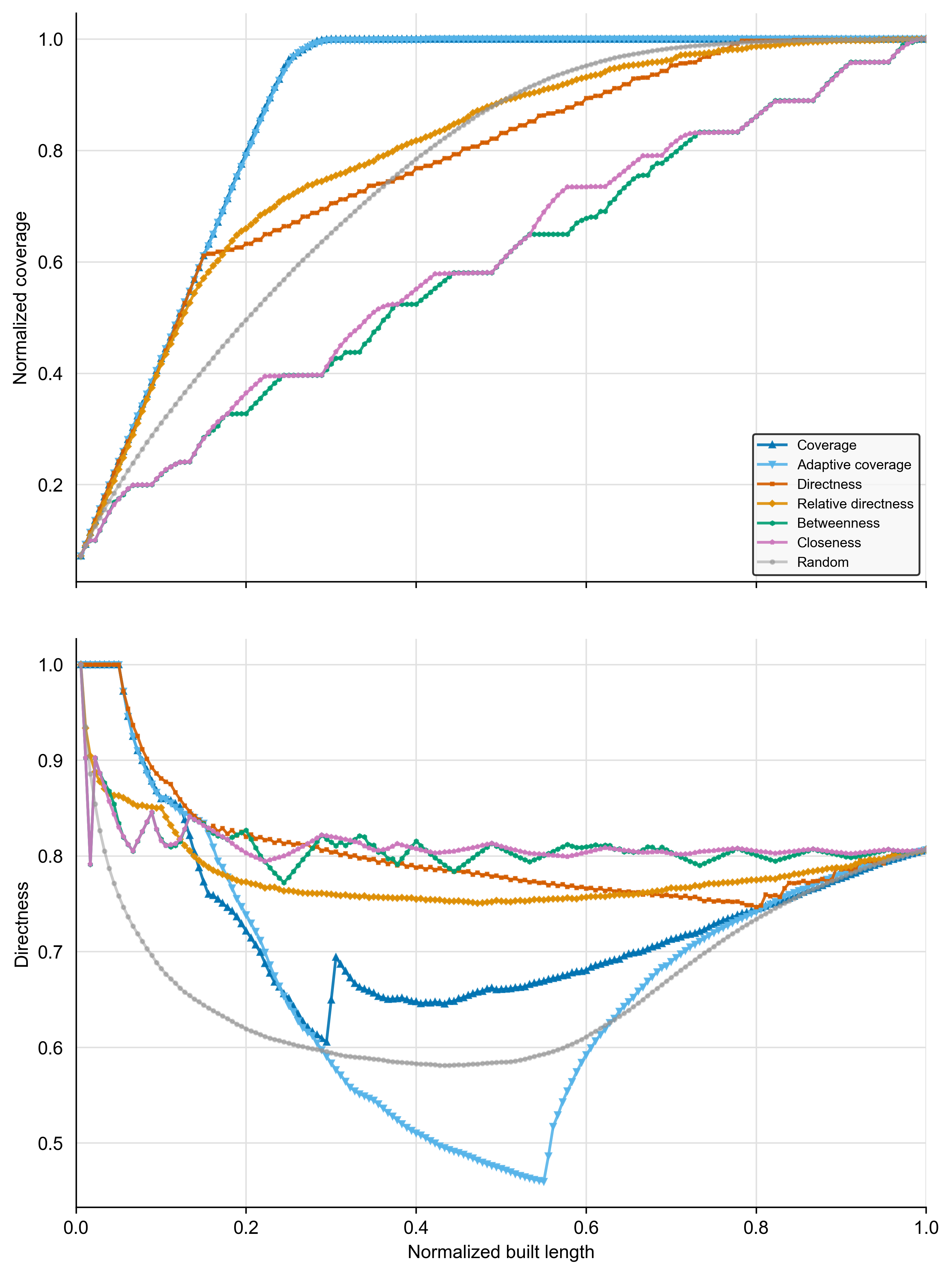}
    \caption{\textbf{Performance metrics per built length.} Average coverage (top) and directness (bottom) per built length for all additive growth strategies, except manual strategies. Coverage is monotonically non-decreasing, while directness can fluctuate heavily due to its non-linear nature. Both coverage and built length are normalized by their final value.}
    \label{fig:cov-dir-lineplot}
\end{figure}

\begin{figure}
    \centering
    \includegraphics[width=\linewidth]{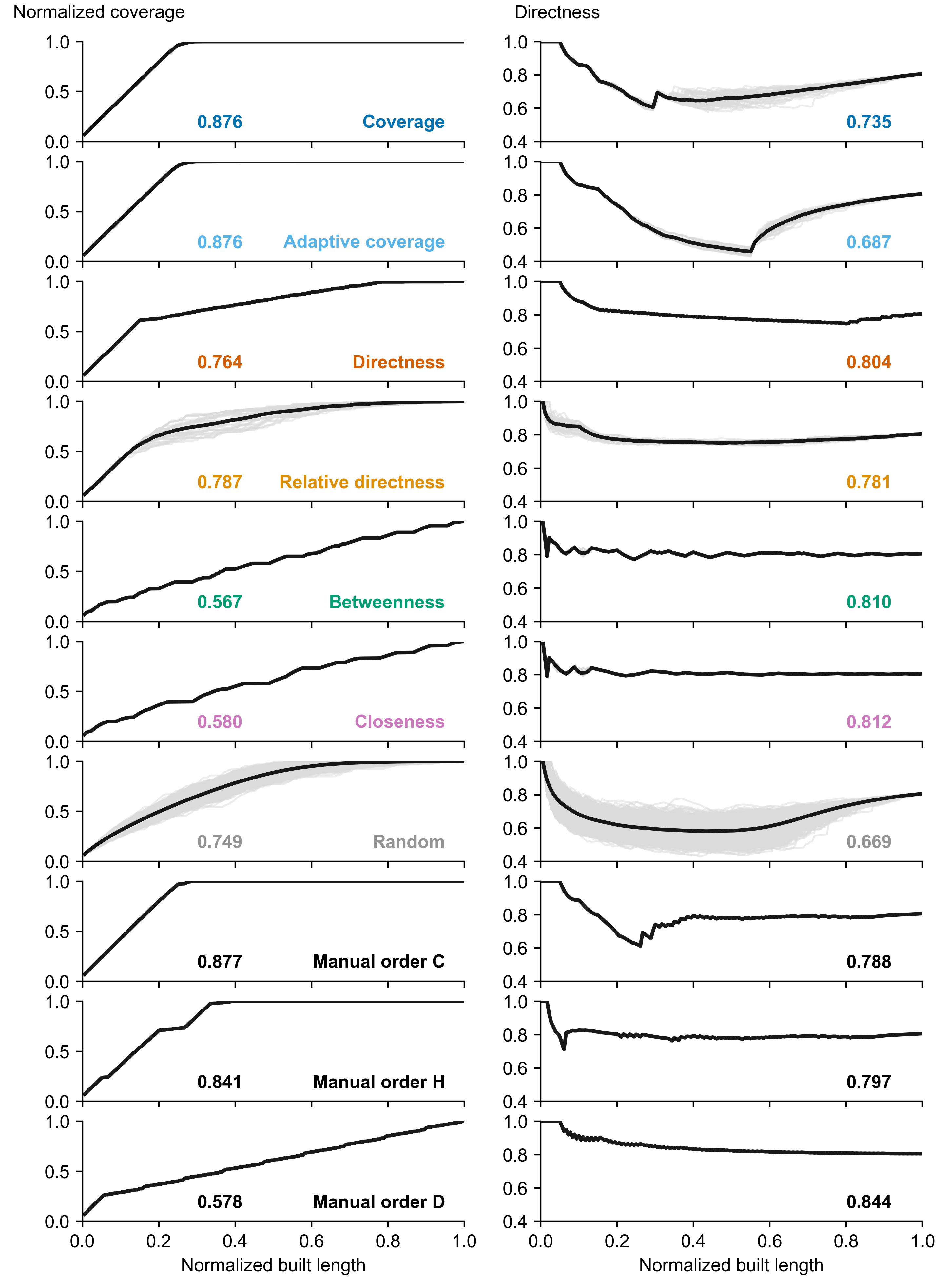}
    \caption{\textbf{Coverage (left column) and directness (right column) versus built length for each growth strategy (rows)}. All 1000 runs are shown in gray and average curves are highlighted in black. Inset values report the areas under the curve (AUC).}
    \label{small-multiples}
\end{figure}

As the network has multiple axes of symmetry and all links have the same length, there can be steps in the greedy optimization process when several links are equally desirable, and the next link is chosen randomly among these equally good options. To observe the impact of these random choices, we run each greedy strategy 50 times. For the random strategy, we perform 1000 runs, as this strategy can theoretically give results from the entire solution space. We report the average performance of each strategy on coverage and directness in Fig.~\ref{fig:cov-dir-lineplot}\textbf{a} and \textbf{b}, respectively. Figure~\ref{small-multiples} visualizes the curves from Fig.~\ref{fig:cov-dir-lineplot} as small multiples, while additionally reporting the outcomes of all single runs (gray curves). We set our length unit to kilometers due to the most intuitive correspondence with transport network applications, but this choice is arbitrary.

\begin{figure}
    \centering
    \includegraphics[width=0.75\linewidth]{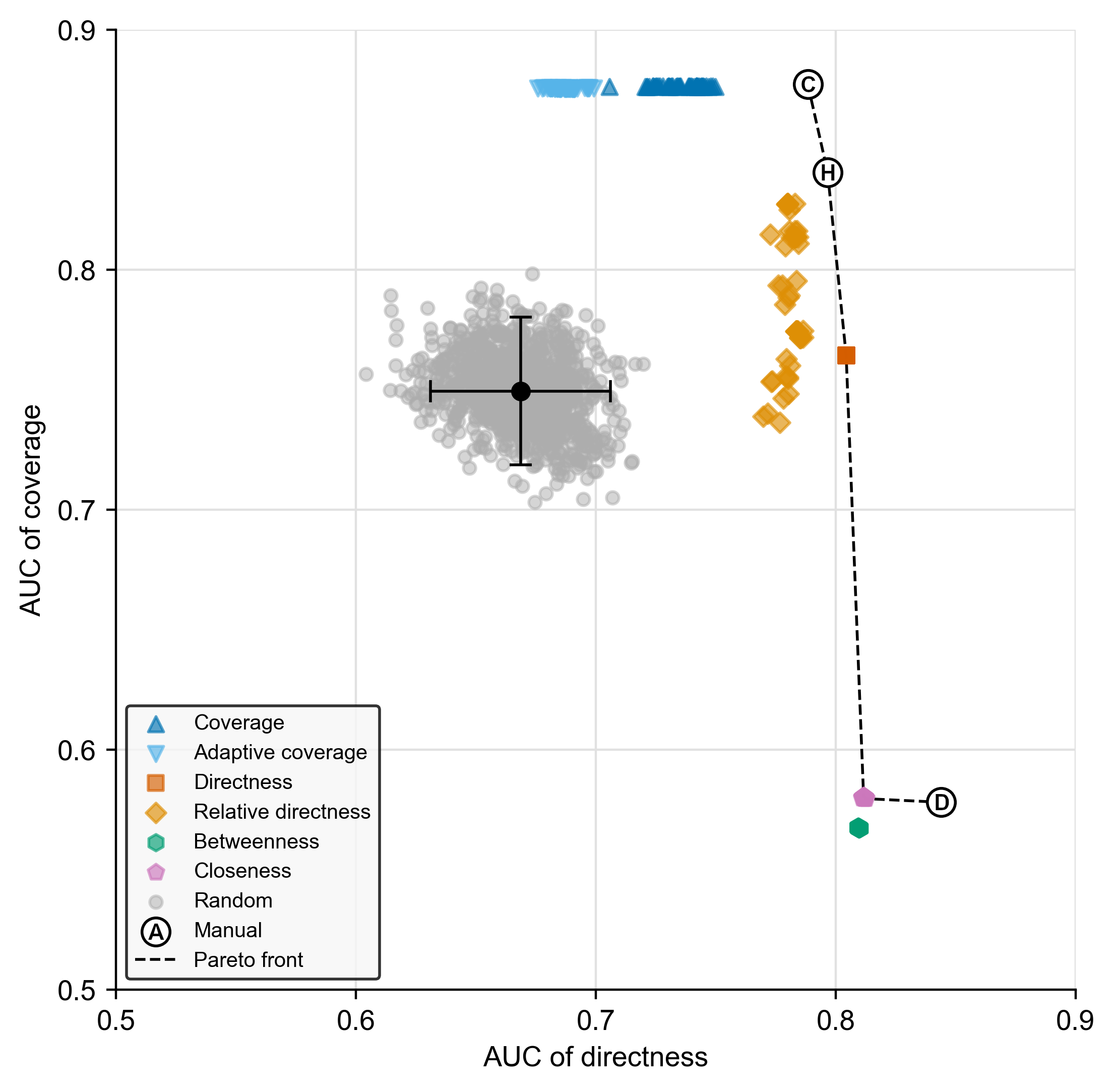}
    \caption{\textbf{AUC of coverage against AUC of directness for all additive growth strategies.} The black dot is the mean Random strategy over 1000 runs, each error bar depicts two standard deviations. The random orders are over six standard deviations away from the Pareto front, showing that brute-forcing is infeasible. Coverage and Adaptive coverage yield maximum AUC of coverage, but so is the manual order C which also outperforms both in AUC of directness. The Pareto front is made up of all the manual strategies and Directness and Closeness. Betweenness and Relative directness are slightly sub-optimal.}
    \label{fig:AUC-cov-dir-add}
\end{figure}

Each of our two performance metrics (coverage and directness) provides for each strategy a value for each growth step, yielding a curve over all growth steps. As we aim to quantify the performance across the entire growth process, the Area Under the Curve (AUC) is an appropriate tool. Since the final network is pre-defined, the initial and final values of all curves are known, while the interesting variations happen during intermediate stages of growth. While it is possible to visually compare curves, as in Figs.~\ref{fig:cov-dir-lineplot} and \ref{small-multiples}, the AUC summarizes a whole curve with a single value, and thus allows comparison between multiple strategies in coverage-directness space, see Fig.~\ref{fig:AUC-cov-dir-add}. 

AUC performance metrics are normalized in the range $[0, 1]$. To do so, we first normalize the built length and coverage values by min-max normalization, i.e., we compute the normalized values $x'$ from the original values $x$ using:

\begin{equation}
    x'=\frac{x-\text{min}(x)}{\text{max}(x)-\text{min}(x)}
\end{equation}

We do not normalize directness, as it is by definition bounded within $[0, 1]$. As the initial and final values of our performance metrics are fixed, we normalize the computed AUC for our normalized variables by the AUC obtained with the optimal curve for our performance metrics, i.e., a curve starting at the given initial value, reaching the maximal value for all intermediate stages, and ending at the final value.

The AUC tool, as applied above, weights all growth steps equally. Since it could be argued that earlier growth steps are more important, we explore in Appendix~\ref{exp-disc} also a variant of the AUC with exponential discounting to give more value to earlier growth stages. This variant leads to similar results.

We report our main observations in the following subsections.

\subsection{Greedy strategies outperform random strategy}
\label{rand-out-greed}

We find that typical random orders are highly sub-optimal as their AUC values (cloud of 1000 gray markers in Fig.~\ref{fig:AUC-cov-dir-add}) are far and well separated from the Pareto front (computed for all considered orders, dashed line). In fact, quantifying the distance of the AUC values from the Pareto front, we find that the average random order (black dot) is at least six standard deviations away from the front for each axis (error bars in the black cross denote two standard deviations). Assuming normality, which is justified at least for coverage (Shapiro-Wilk test does not reject at $p=0.89$), it means that on average the Random strategy would need to be run at least 500 million times before finding one order close to the Pareto front. This result shows that, although in theory a random ordering could recreate any other order, most of the possible solution space -- especially high performing solutions -- is unlikely to be reached by brute force. Therefore, at least for the grid, if the aim is to reach a good level of directness or coverage (or both), it is clear that smarter strategies are needed.

The AUC of directness spreads slightly more than the AUC of coverage for random orders. Further, as visible in Figs.~\ref{fig:cov-dir-lineplot} and \ref{small-multiples}, while coverage increases monotonously and by a maximum possible value, Eq.~(\ref{eq:bufferedlink}), directness is non-linear and can feature long jumps triggered by the addition of one single link. These observations give a first hint that optimizing for directness is harder (see Section~\ref{dir-local-trap}).

Concerning the spread of the greedy strategies, three of them are deterministic (Directness, Closeness, and Betweenness), leading just one marker being visible for each in Fig.~\ref{fig:AUC-cov-dir-add} because the 50 runs produce the same AUC values. The three others (Coverage, Adaptive coverage, and Relative directness) involve a strong random element leading to scattered AUC values from their 50 runs. The Coverage and Adaptive coverage orders (blue triangle markers) reach the same AUC of coverage but different AUC of directness, with Coverage always outperforming Adaptive coverage (all dark blue markers are further right than the light blue markers). Relative directness orders (orange diamond markers) are spread along both dimensions but more vertically, along AUC of coverage. The Directness strategy (red square marker) always reaches higher AUC of directness but most Relative directness orders reach higher AUC of coverage. Interestingly, both the Closeness and Betweenness strategies reach slightly higher AUC of directness than the Directness strategy (albeit with much lower AUC of coverage), implying that the Directness strategy is not globally optimal. We give a more detailed explanation of the scattering of AUC values in Section~\ref{dir-local-trap}.

\subsection{Manual orders outperform greedy strategies}

When accounting for manual orders, we first find that the only greedy orders on the Pareto Front are Directness and Closeness (red square and purple pentagram markers in Fig.~\ref{fig:AUC-cov-dir-add}, respectively). The Directness strategy has slightly more AUC of directness than manual order H, while the Closeness strategy has slightly more AUC of coverage than the manual order D. 

The manual orders C and D lie on the Pareto Front, which was expected since we designed them to be optimal strategies. Manual order H, despite only following a hierarchical heuristic, is also on the Pareto Front, showing the performance of a simple intuitive heuristic.

The manual order C is on the same maximal level of AUC of coverage as the Coverage or Adaptive coverage orders, but it beats all of them in terms of AUC of directness. This difference in directness is caused by the randomness of choices made by the greedy strategies when full coverage is reached (see Section~\ref{cov-limited-random}).

The manual order D has much worse AUC of coverage than all the Directness and Relative directness orders, but outperforms them in terms of AUC of directness. This gap can be explained by the local optimal traps a greedy algorithm can fall into (see Section~\ref{dir-local-trap}).

\subsection{Coverage strategies reach full coverage, then become random}
\label{cov-limited-random}

When growing a planar network with straight links, eventually a point is reached where no additional link adds more coverage. We call this point ``reaching full coverage''. The number of links needed to reach full coverage depends on the buffer size. Let us consider our $10\times10$ grid with 180 links, each 100 meters long. 

If the buffer size is at most 50 meters, all 180 links ($100\%$) are needed to reach full coverage, since each new link will add coverage, no matter how many previous links there are. In the other extreme, if the buffer size is at least 450 meters, only the 36 links ($20\%$) on the periphery are needed to reach full coverage, since links on the inside would not add any extra coverage. If the buffer size is intermediate, the network reaches full coverage with an intermediate number of links. 

\begin{figure}[t]
    \centering
    \includegraphics[width=0.6666\textwidth]{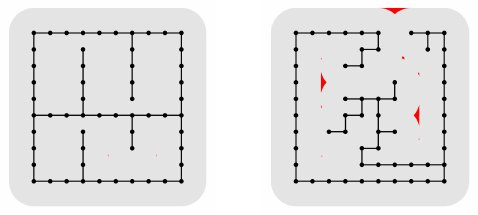}
    \caption{\textbf{Coverage strategy snapshot illustrating the effect of randomness.} Left: additive, Right: subtractive. Shown are two typical networks at step 58. The additive strategy has almost reached full coverage. At the same stage, the subtractive strategy is non-deterministic and has reached a meandering form that is underperforming in both coverage and directness. Highlighted in red is the difference to the full coverage.}
    \label{fig:coverage-order}
\end{figure}

In the case of the selected intermediate buffer size of 150 meters, with a single connected component the full coverage can be reached with 48 links ($27\%$), which is achieved by the manual order C, see central panel of Fig.~\ref{fig:manual}. Starting from the link with largest average closeness, 58 links ($32\%$) are necessary to reach full coverage, which is achieved by the additive Coverage strategy, see left panel of Fig.~\ref{fig:coverage-order}.

After reaching full coverage, the remaining links are equivalent when optimizing for coverage. Thus, every link added after the first 59 is chosen at random. This random choice results in additive Coverage orders sharing the same curve for coverage, but varying in directness after reaching the full coverage at $33\%$ of total length built (see Fig.~\ref{small-multiples}). In contrast, manual order C follows the hierarchical order H (see Section~\ref{man-hier-ord} and \ref{man-cov-ord}) after having reached full coverage optimally. By switching the objective here, manual order C manages to be well-performing on both directness and coverage.

Let us now consider what happens with a subtractive strategy. As explained in Section~\ref{greedy-optim}, subtractive strategies start from the final network, removing the link of the least importance for the chosen metric. When optimizing for coverage however, any link on the inside of the network removes the same amount of coverage and is thus equivalently ``least important''. Thus, the first links removed are random links from the inside. The remaining reverse growth process, while optimizing for coverage, follows highly constrained, typically meandering shapes (see right panel of Fig.~\ref{fig:coverage-order}). These shapes have worse directness than in the additive Coverage strategy (see left panel of Fig.~\ref{fig:coverage-order}), leading to the subtractive Coverage strategy underperforming on both coverage and directness, compared to the additive Coverage strategy (see Section~\ref{add-vs-sub}).

\subsection{Directness falls into a local optimum trap}
\label{dir-local-trap}

\begin{figure}[t]
    \centering
    \includegraphics[width=0.6666\textwidth]{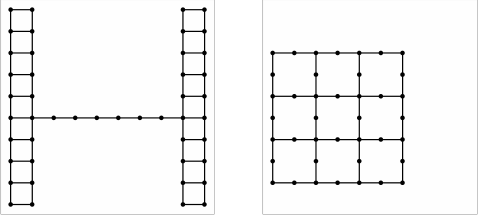}
    \caption{\textbf{Directness strategy snapshots illustrating local optimum traps.} Left: additive, step 63. Right: subtractive, step 48. The additive strategy avoids building cycles, but if it has to, they will be as short as possible. On the contrary, the subtractive strategy avoids breaking up cycles, leading to large cycles constructed first.}
    \label{fig:directness-local-trap}
\end{figure}

\begin{figure}[t]
    \centering
    \includegraphics[width=0.6666\textwidth]{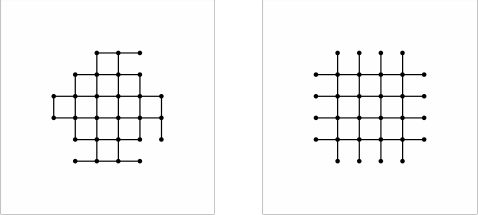}
    \caption{\textbf{Centrality-based strategy snapshots illustrating ``well-behaved'' growth.} Network at step 40 with the Betweenness (left) and Closeness (right) strategies. Both growth processes create small cycles from the center, which keeps directness relatively high and stable (at the cost of low coverage).}
    \label{fig:bet-vs-clo}
\end{figure}

As directness is heavily non-linear, locally optimal choices can easily trap the greedy optimization process. Cycles are a good example to understand why, as shown in Fig.~\ref{fig:directness-local-trap}. With a fixed link length, the longer the cycle (i.e., the more links it contains), the lower is its directness. Therefore, when the Directness strategy is additive, cycles are only created whenever they are one step away to be completed -- a state that is increasingly harder to reach the longer the cycle is, see left panel of Fig.~\ref{fig:directness-local-trap}. However, when the Directness strategy is subtractive, the longer the cycle, the bigger the impact of removing one single link, which can result in holding on to cycles for too long. In other words, the subtractive algorithm grows large cycles first. Both additive and subtractive issues are undesirable behaviors stemming from local optimum traps for directness.

Local optimum traps also explain why strategies based on centrality (i.e., Betweenness and Closeness) perform in directness better than the Directness strategy itself. As the centrality value of each link is fixed by the final network, centrality-based strategies follow a simple heuristic, creating small cycles from the center, see Fig. \ref{fig:bet-vs-clo}. They keep a value close to the final directness throughout the growth process, with some oscillations, see the green and pink curves in the bottom panel of Fig.~\ref{fig:cov-dir-lineplot}. On the other hand, the Directness strategy locks itself into a local optimum trap after constructing a straight line of maximal directness, see left panel of Fig.~\ref{fig:directness-local-trap}. This trapping leads to a slow but steady decrease of the directness for most of the growth process, see the orange curve in the bottom panel of Fig.~\ref{fig:cov-dir-lineplot}. Putting higher value on early growth stages via exponential discounting (see Appendix~\ref{exp-disc}) makes the Directness strategy perform better, but does not change the relative position of most strategies overall.

While we can easily optimize to get the best or worst AUC of coverage possible using greedy growth, we cannot do the same for AUC of directness due to its non-linear nature. However, a greedy process could account for the 4-cycles in a grid lattice by increasing the time horizon from one to three future steps (see Section~\ref{other-toy-net} for other examples).

\subsection{Coverage vs.~adaptive coverage, directness vs.~relative directness}
\label{var-can-met}

In this section, we explore the variants of coverage (adaptive coverage) and directness (relative directness). 

\begin{figure}[t]
    \centering
    \includegraphics[width=0.9999\textwidth]{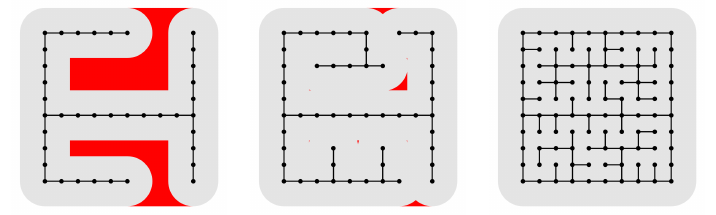}
    \caption{\textbf{Adaptive coverage strategy snapshots.} From left to right: network at step 37, 51, and 101, respectively. Full coverage is reached only at step 101. Highlighted in red is the difference to the full coverage.}
    \label{fig:adap-cov-viz}
\end{figure}

Adaptive coverage orders are worse than Coverage orders for both AUC of coverage and directness, compare light blue with dark blue markers in Fig.~\ref{fig:AUC-cov-dir-add}. To measure coverage, we choose a single buffer size. The area measured is dependent on both the network structure, and the selected buffer size. Since the Adaptive coverage strategy optimizes across different scales through growth via a buffer size change, we expect its AUC of coverage -- which is measured for one single buffer size -- to be worse than the AUC of coverage for the Coverage strategy which only optimizes for that single buffer size (see Fig. \ref{fig:cov-dir-lineplot}). Both strategies optimize by covering the greatest area possible, which for a lattice results in a sparse network of high coverage but low directness. While the Coverage strategy reaches full coverage with 59 links, see left panel of Fig.~\ref{fig:coverage-order}, the Adaptive coverage strategy does so only when it reaches 101 links, with the buffer size decreasing at 37 and 51 links, see Fig.~\ref{fig:adap-cov-viz}. After reaching full coverage, 70 links are randomly added in the Adaptive coverage strategy versus 121 links in the Coverage strategy. Consequently, the decrease of directness lasts longer for the Adaptive coverage orders, see Fig.~\ref{fig:cov-dir-lineplot}.

\begin{figure}
    \centering
    \includegraphics[width=0.6666\textwidth]{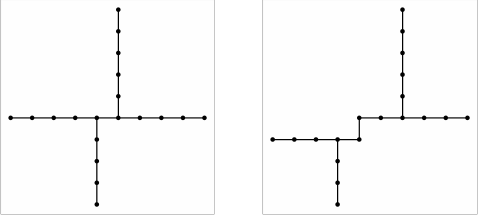}
    \caption{\textbf{Relative directness strategy snapshots illustrating random variations.} Two examples of a network at step 18 having maximal relative directness.}
    \label{rel-dir-examples}
\end{figure}

We defined previously relative directness as the directness where the Euclidean distance is replaced by the final shortest network path length. Given that we work on a grid, the final shortest network path between any node pair will be a sequence of horizontal and vertical edges, implying that the path's length is then equivalent to the Manhattan distance. Consequently, maximal relative directness is achieved by different link orderings with varying AUC of coverage, see Figs.~\ref{rel-dir-examples} and \ref{small-multiples}.

\subsection{Additive strategies outperform subtractive strategies}
\label{add-vs-sub}

As the subtractive strategies start with more information on the network, one could expect these strategies to outperform the additive ones. However, when taking together additive and subtractive greedy strategies and comparing them, we find that no subtractive strategy comes to lie on the Pareto front for the grid, except centrality-based strategies that are equivalent in both orders, see Fig.~\ref{AUC-sm}.

The lower performance of subtractive strategies can be explained by already noted behaviors: As optimizing for coverage after having reached maximal coverage implies random link choice, subtractive Coverage and Adaptive coverage strategies start randomly. The optimization phase is then heavily constrained by the random configuration of maximal coverage obtained. On the other hand, additive Coverage and Adaptive coverage strategies start by efficiently covering the area, creating a more optimal growth, see Fig.~\ref{fig:coverage-order}.

The additive Directness strategy starts with a straight line along the first link, then creates perpendicular straight lines at the extremities, from where small cycles are created. Subtractive Directness strategies, on the other hand, can remove different links that are all producing subnetworks of equal directness, creating the variability in AUC of coverage. From those multiple configurations typically emerge large cycles that are harder to break, see Fig.~\ref{fig:directness-local-trap}. While both behaviors are suboptimal, additive strategies outperform subtractive ones.

Additive strategies outperform subtractive ones on all tested networks, except for the subtractive Adaptive coverage and Coverage strategies on the diagonal grid and the radio-concentric networks. The difference can be small: On the grid, the highest AUC of coverage for the additive and subtractive Coverage strategies is 0.87647 and 0.87633, respectively, and the highest AUC of directness for Directness strategies is 0.80438 versus 0.80157. The difference between additive and subtractive strategies is much smaller than the difference between random and greedy strategies. Subtractive strategies can perform better for specific conditions, depending on the chosen performance metric and the network structure.

\subsection{Results for other stylized networks are similar}
\label{other-toy-net}

\begin{figure}
    \centering
    \includegraphics[width=0.9999\textwidth]{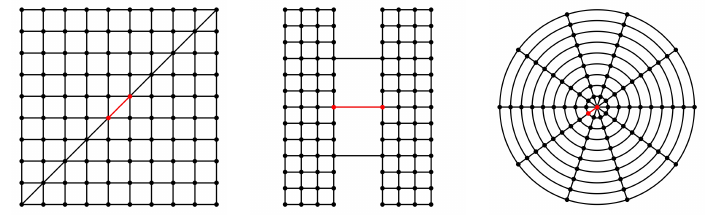}
    \caption{\textbf{Additional tested stylized networks.} Left: Grid with a diagonal, Center: Three bridges network, Right: Radio-concentric network. Highlighted in red are the initial link and nodes for the additive strategies.}
    \label{toy-graph-fig}
\end{figure}

Expanding our analysis, we run the random and greedy strategies on three other stylized networks, see Fig. \ref{toy-graph-fig}. All networks have similar size, around 100 nodes and 180 links. The selection of stylized networks follows typical urban street patterns, explained in Appendix \ref{urbantoynet}.

\subsubsection*{Grid with a diagonal}

The ``Grid with a diagonal'' network is a $10\times10$ grid with links along one diagonal, resulting in 189 links. Links along the diagonal are $\sqrt{2}$ times longer than the other links. This network is the most similar to the regular grid, with the diagonal acting both as added asymmetry and a “shortcut” through the network. The initial link is at the center of the diagonal. 

\begin{figure}
    \centering
    \includegraphics[width=0.6666\textwidth]{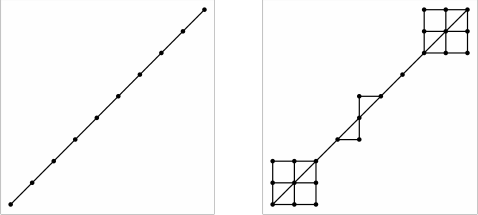}
    \caption{\textbf{Grid with a diagonal illustrating the Directness strategy closing 3-cycles early.} Left: step 9, Right: step 37.}
    \label{gd-dir}
\end{figure}

In contrast to to the regular grid, the AUC of directness of all growth strategies is higher for the Grid with a diagonal for almost all growth strategies (except Coverage and Adaptive coverage), see Fig.~\ref{AUC-sm}~\textbf{c}. This can be explained by the higher directness of the final network ($D \approx 0.81$ for the grid versus $D \approx 0.85$ for the grid with a diagonal), as the Grid with a diagonal has nine additional links compared to the grid. The Directness strategy outperforms all other strategies on directness. After growing the diagonal, one adjacent link is added. Since cycles adjacent to the diagonal cycles contain three (rather than four) links, the Directness strategy can find these 3-cycles and create them throughout the growth process, in this manner avoiding some local optimum traps, see Fig.~\ref{gd-dir}. 

\begin{figure}
    \centering
    \includegraphics[width=0.6666\textwidth]{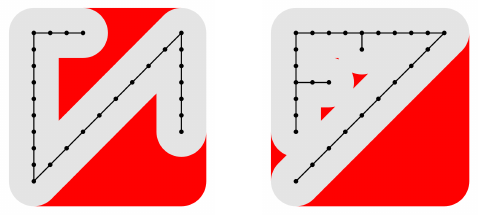}
    \caption{\textbf{Grid with a diagonal illustrating the Adaptive coverage slightly outperforming the non-adaptive Coverage strategy in terms of coverage.} Left: Adaptive coverage strategy, covering $68.2\%$ of the full coverage. Right: non-adaptive Coverage strategy, covering $65.9\%$ of the full coverage. Both at step 27. Highlighted in red is the difference to the full coverage.}
    \label{gd-cov-ad}
\end{figure}

The Adaptive coverage strategy slightly outperforms the non-adaptive Coverage strategy on coverage. Since the starting link is on the diagonal, both these growth strategies first construct all links along the diagonal. Thereafter, the Adaptive coverage strategy starts building the periphery, because for a larger buffer size these are the only edges that improve the coverage, see left panel of Fig.~\ref{gd-cov-ad}. However, for the non-adaptive Coverage strategy, the links on the periphery adjacent to the diagonal act as a local optimum trap due to the small buffer size. Building a new side for the periphery to grow is more costly than continuing growth on a single side of the periphery. This behavior leads the non-adaptive Coverage strategy to delay constructing the periphery, see right panel of Fig.~\ref{gd-cov-ad}.

\subsubsection*{Three bridges}

The ``Three bridges'' network consists of two $4\times13$ nodes grids connected by three links $3$ times longer than the other links, resulting in 104 nodes and 177 links. While most of the network is still grid-like, the three bridges are critical links of high value for directness, coverage, and the shortest path affecting the centrality-based metrics. The initial link is the central bridge.

\begin{figure}
    \centering
    \includegraphics[width=0.6666\textwidth]{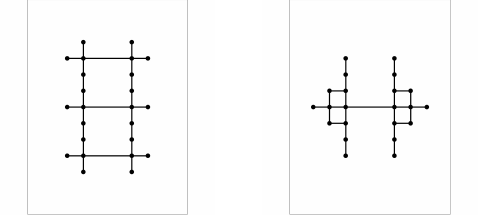}
    \caption{\textbf{All three bridges are built early with the Betweenness but not with the Closeness strategy.} Left: Betweenness strategy, Right: Closeness strategy. Both at step 25.}
    \label{tb-fig:bet-vs-clo}
\end{figure}

Obtained results are similar to the regular grid, see Fig.~\ref{AUC-sm}~\textbf{e}. The gap in AUC of coverage between the Coverage and Adaptive coverage orders and some other orders is larger. Since the three bridge links contribute a significant share to the total coverage, random orders have lower AUC of coverage. While centralities are limited by the growth diffusing from the center, the bridge links have high betweenness, thus they are more quickly built by the Betweenness strategy than by the Closeness strategy, see Fig.~\ref{tb-fig:bet-vs-clo}. As the Directness strategy grows in a similar manner as the regular grid, see Fig.~\ref{fig:directness-local-trap}, bridges are added at the latest stages. The consequences are compensated for by the reduced amount of connections between the extremities of the network making shortest paths longer, and thus lowering the overall directness ($D \approx 0.8064$ for the regular grid, while $D \approx 0.7923$ for the Three bridges network).

\subsubsection*{Radio-concentric}

The ``Radio-concentric'' network is made up of 10 radial lines intersecting with 9 circles, resulting in 91 nodes and 180 links. Links along the radial lines are all 100 meters long, while links along the circles are of length $2\pi k/10$, with $k$ numbering the circles from the center ranging from $1$ for the smallest to $9$ for the largest, from around 63 meters to 565 meters. The initial link is inside the first circle, along a radial line.

\begin{figure}
    \centering
    \includegraphics[width=0.9999\textwidth]{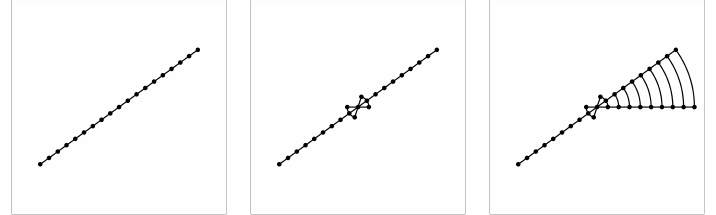}
    \caption{\textbf{The Radio-concentric network builds up radial links first with the Directness strategy.} Left: step 18, Center: step 26, Right: step 42.}
    \label{rc-dir}
\end{figure}

Obtained results are similar to the regular grid, see Fig.~\ref{AUC-sm}~\textbf{g}. The range of values of AUC of directness is larger though, reaching less than 0.6 for random strategies and over 0.9 for some non-random strategies. This larger range is caused by multiple factors. First, the higher ratio of links per nodes increases the directness of the final network ($D \approx 0.8064$ for the regular grid, while $D \approx 0.8930$ for the Radio-concentric network). Second, shortest paths are less evenly distributed than on the grid. Third, the first circle is made up of three-link cycles that are easier to grow for the Directness strategy, as explained in Section~\ref{dir-local-trap} (see Fig.~\ref{rc-dir}). Finally, angles between links from different radial lines are smaller than the $90^{\circ}$ angle of the grid, reducing the gap between Euclidean distances and shortest network paths.

\subsection{Robustness to exponential discounting and distortions}
See Appendix~\ref{exp-disc} for a variant of the AUC where we use exponential discounting to give more value to earlier growth stages, leading to similar results. All results are also robust to distortions of the grid, except for Centrality strategies, see Appendix~\ref{robust-distortion}.

\section{Discussion and conclusion}\label{sec:discussion}

Our study explored systematically the growth of spatial networks and their performance in directness-coverage space, to provide fundamental insights into the trade-offs between these two key metrics in network design applications. We focused on transport applications, but our study could be relevant for other applications that involve spatial networks. Motivated by the transport perspective, we focused on the case of a given set of nodes with a final set of links between them that could be built over time, and on the order in which these links are added. We pursued and compared the idiosyncrasies of several strategies down to specific network configurations trapped in local optima, involving random growth, greedy optimization of several metrics, and manual designs.

Overall, we confirm the intuitive expectation that greedy growth is generally outperforming random growth, and that well-designed manual growth strategies can surpass greedy strategies, even those that locally optimize for the global target metric. The main added benefits of our study is 1) \emph{a quantification} of how much worse such strategies are, and 2) \emph{an ordering} of several metrics in the directness-coverage space, showing which of them are worth pursuing through greedy optimization for which goal. For example, on a grid-like city plan, Directness always outperforms Relative directness, and Coverage always outperforms Adaptive coverage, in terms of being closer to the Pareto front.

In a practical design process, our study shows that, for example, it can make sense to first generate a link ordering based on greedy coverage optimization towards global coverage optimization, to be then taken as a starting point for a manual tuning to achieve higher directness. Holding for a number of stylized urban transport network topologies, we found that centrality-based greedy strategies tend to perform best for directness but are worse than random strategies for coverage, while coverage-based greedy strategies can achieve maximum global coverage as fast as possible but perform similarly bad for directness as random strategies.

Since the solution space is exponentially large and thus intractable, we quantify how greedy optimization -- despite usually not providing the globally optimal solution -- is giving much better results than random trials. In fact we quantify the large extent to which brute-forcing is infeasible already for small networks, where over 500 million random trials would be needed to find a solution close to the Pareto front. Techniques such as simulated annealing or evolutionary algorithms \citep{martinez2016cycling} could yield better solutions than our greedy algorithms and could be a future algorithmic improvement. A further improvement could be the creation of a step-dependent optimization function, for example switching from a greedy coverage strategy to a greedy directness strategy. Such a switch could be justified if there are different objectives at different stages of the growth process.

The intuitive insight of ``manual beats greedy beats random'' being confirmed in our context has practical consequences. In practice, the growth of transport networks can indeed be close to random or greedy. This feature is most prevalent in networks where the cost or friction of adding links is high and where single links are perceived as negligible for network overall function, such as in bicycle networks \citep{szellGrowingUrbanBicycle2022}. Here, the addition of many single bicycle lanes can be seen as a politically contested competition for car parking space \citep{oldenziel2011contested,gossling2020cities}. Further, political competences can be fragmented: for example, a city might have an all-encompassing bicycle network plan in place, but individual districts may ignore it. We are aware of anecdotal evidence for this effect in cities like Paris, Vienna, or for Copenhagen versus Frederiksberg \citep{carstensen_spatio-temporal_2015,paulsen_societally_2023}. Due to such effects, the addition of links proceeds by lowest political friction, which in network terms is indistinguishable from a random growth strategy \citep{szellGrowingUrbanBicycle2022}, since these local decisions are not accounting for the wider network structure, thus do not follow a greater vision. 

Short-term horizons further appear naturally in the context of bicycle networks, where urban or regional planners might aim to implement a time-limited plan for a specific neighborhood or region \citep{rich2021cost,paulsen_societally_2023} (i.e., locally optimizing coverage and directness), or to close certain gaps \citep{nateraorozcoDatadrivenStrategiesOptimal2020,vybornova_automated_2022} (i.e., locally optimizing for connectedness and directness). Even when city-wide ``master plans'' exist, there can be long delays in their realization \citep{flyvbjerg2005accurate,carstensen_spatio-temporal_2015}. Our results show that in general it is better to implement a global plan than to follow locally or randomly motivated growth. At the same time, we must be aware of the long history of failed ``central plans'' in urban planning \citep{batty2013new}, thus practical feasibility and a natural bottom-up component of city development must be accounted for.

One further notable insight of our study is that additive strategies generally outperform subtractive strategies. This insight gives an unexpected contrasting result to four recent studies on bicycle network design \citep{steinacker_demand-driven_2022,ballo2024designing,steinacker2025robust,wiedemannBikeNetworkPlanning2025} that argue in favor of the subtractive ``backwards percolation'' method. Although our greedy single metric optimization is not directly comparable to the greedy optimization of the bikeability metrics used in those studies, our result calls for care when deciding between additive and subtractive methods: both should always be tested before settling for one of them.

Our last discussion point is on realism. Our study purposefully analyzes network growth on stylized topologies only, in order to provide a \emph{systematic} insight into the trade-offs between directness and coverage. We also use metrics like betweenness as a crude proxy for flow, while in reality there is an underlying population distribution and origin-destination matrix, where travelers, especially cyclists, might not take shortest paths but could be strongly influenced by the presence of slopes, green and blue spaces, or vehicular traffic. As such, this research is highly reductionist and fundamental, and thus in itself not directly applicable to practical network design problems. Nevertheless, our results have profound implications for the general design of spatial networks, especially planar transport networks, and provide different starting points for planners who aim to achieve higher performance in specific network metrics. As \cite{steinacker2025robust} have recently argued, a network science based approach like ours is justifiable despite being more approximative compared to dynamic cost-benefit approaches from transport planning, due to its qualitatively similar results and ease of implementation. 

The next step following this research should be its application to empirical street networks or real-life (bicycle network) plans. While this is a straightforward call to make, it is also a challenging task as we expect difficulties due to computational complexity. Directness calculations could scale as $O(L^5)$ with $L$ being the number of links, due to the need to recalculate all-to-all shortest paths in each step; and where manual orders would be much harder to design due to the larger size and higher complexity of empirical networks. Nonetheless, we anticipate such computational issues to be eventually resolved, breaking grounds for the application of our framework to real transport networks, and potentially also to unexpected applications to other network design problems such as telecommunication networks or electric grids.

\section*{Declarations}

\subsection*{Availability of data and materials}
\label{code-availability}
This paper does not use any input data. All materials are available online for reproducing figures, results, and generated data, see \cite{github_ob}.

\subsection*{Competing interests}
The authors declare no competing interests.

\subsection*{Funding}
C.S. and M.S. acknowledge support by the European Union through the Horizon Europe grant JUST STREETS (Grant agreement ID: 101104240). A.V. acknowledges support from Villum Fonden through the Villum Young Investigator programme (project number: 00037394).

\subsection*{Author contributions}

\textbf{Clément Sebastiao:} Conceptualization, Formal analysis, Investigation, Methodology, Software, Visualization, Writing – original draft, Writing – review and editing. \textbf{Anastassia Vybornova:} Conceptualization, Writing - review and editing. \textbf{Ane Rahbek Vierø:} Conceptualization, Writing – review and editing. \textbf{Luca Maria Aiello:} Conceptualization, Supervision, Writing – review and editing. \textbf{Michael Szell:} Conceptualization, Funding acquisition, Methodology, Project administration, Supervision, Visualization, Writing – original draft, Writing – review and editing.

\subsection*{Acknowledgements}
Not applicable.

\begin{appendices}

\section{Additional plots}

\begin{figure}[p]
    \centering
    \includegraphics[width=\linewidth]{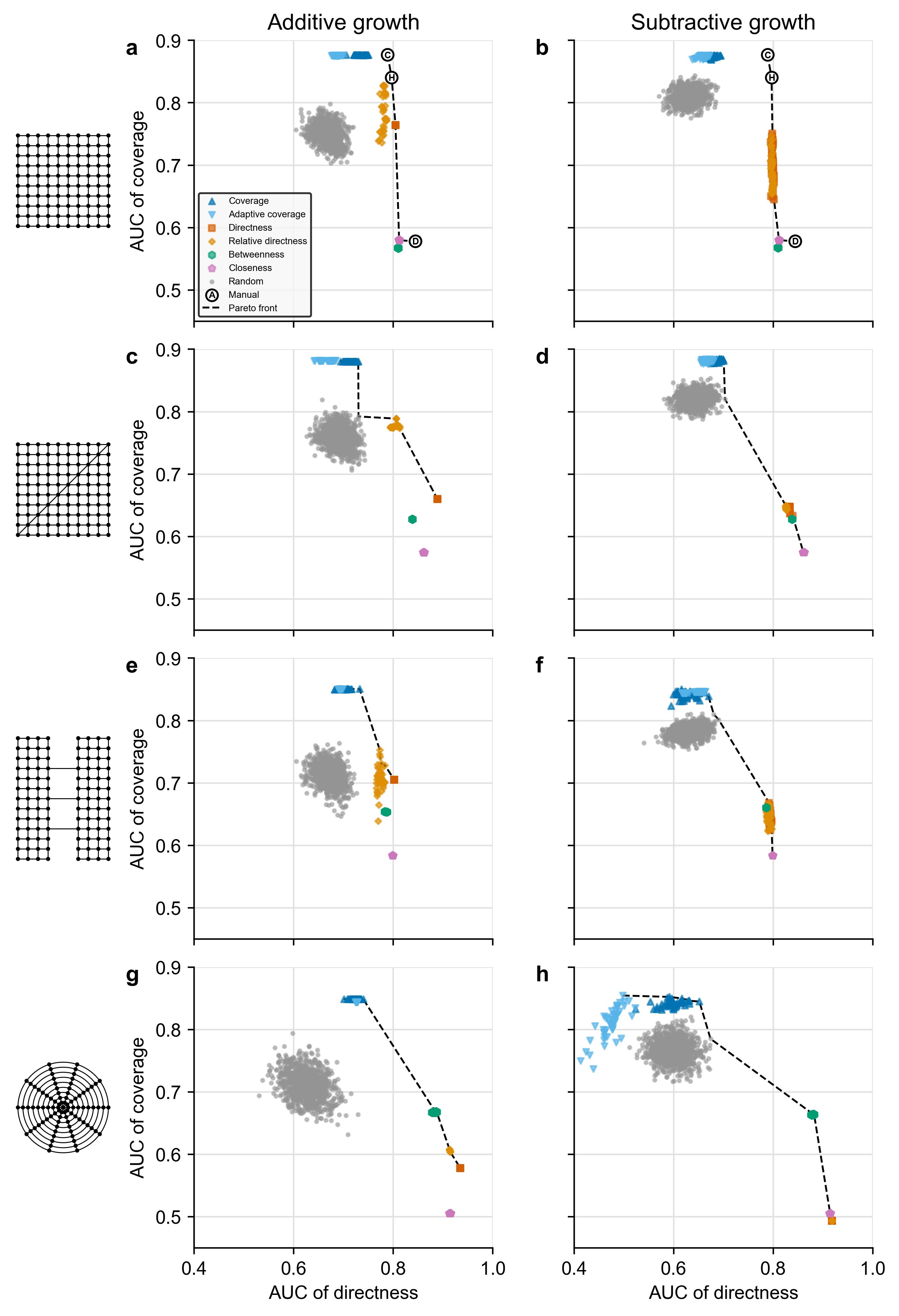}
    \caption{Comparison of the AUC of coverage and directness for \textbf{a} all additive growth strategies on a square grid, \textbf{b} all subtractive growth strategies on a square grid, \textbf{c} all additive growth strategies on a square grid with diagonal, \textbf{d} all subtractive growth strategies on a square grid with diagonal, \textbf{e} all additive growth strategies on a square grid with bridges, \textbf{f} all subtractive growth strategies on a square grid with bridges, \textbf{g} all additive growth strategies on a radio-concentric network, \textbf{h} all subtractive growth strategies on a radio-concentric network.\label{AUC-sm}}
\end{figure}

\clearpage
\section{Results are robust to grid distortion}
\label{robust-distortion}

\begin{figure}
    \centering
    \includegraphics[width=\linewidth]{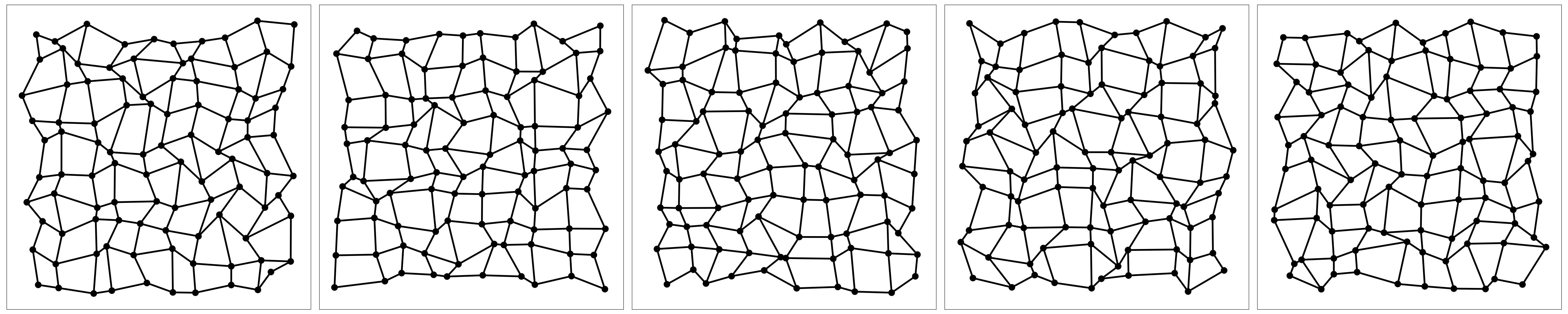}
    \caption{Distorted grids tested.}
    \label{distorted-grids}
\end{figure}

We tested the robustness of our results to grid distortion by re-running the analysis on randomly distorted grids (see Fig.~\ref{distorted-grids}). Grids are distorted by randomly shifting the position of each node by up to $\pm 45$ meters in a uniform distribution. This method is chosen as it ensures that the network stays planar.

\begin{figure}
    \centering
    \includegraphics[width=\textwidth]{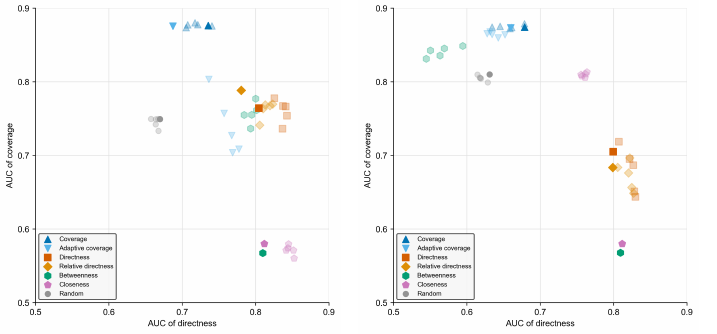}
    \caption{Comparison of the median AUC of coverage and directness for different growth strategies on the grid and distorted grids for the additive (left) and subtractive (right) order.}
    \label{AUC-distorted}
\end{figure}

As is shown in Fig.~\ref{AUC-distorted}, overall, the relative position of all strategies is preserved, except for Centralities and additive Adaptive coverage. We also observe that the directness of all tested distorted grids ($D \in [0.846, 0.858]$) is higher than on the Regular grid lattice ($D \approx 0.806$).

\section{Stylized urban street networks}
\label{urbantoynet}

Since we are interested in the temporal evolution of network growth efficiency, relevant use cases concern network usage during growth in a long enough timescale. The most typical example is the case of an urban transportation network. The completion of bicycle network plans, for example, takes years or even decades to complete in practice, but they are used throughout the entire development period.

\begin{figure}
    \centering
    \includegraphics[width=\linewidth]{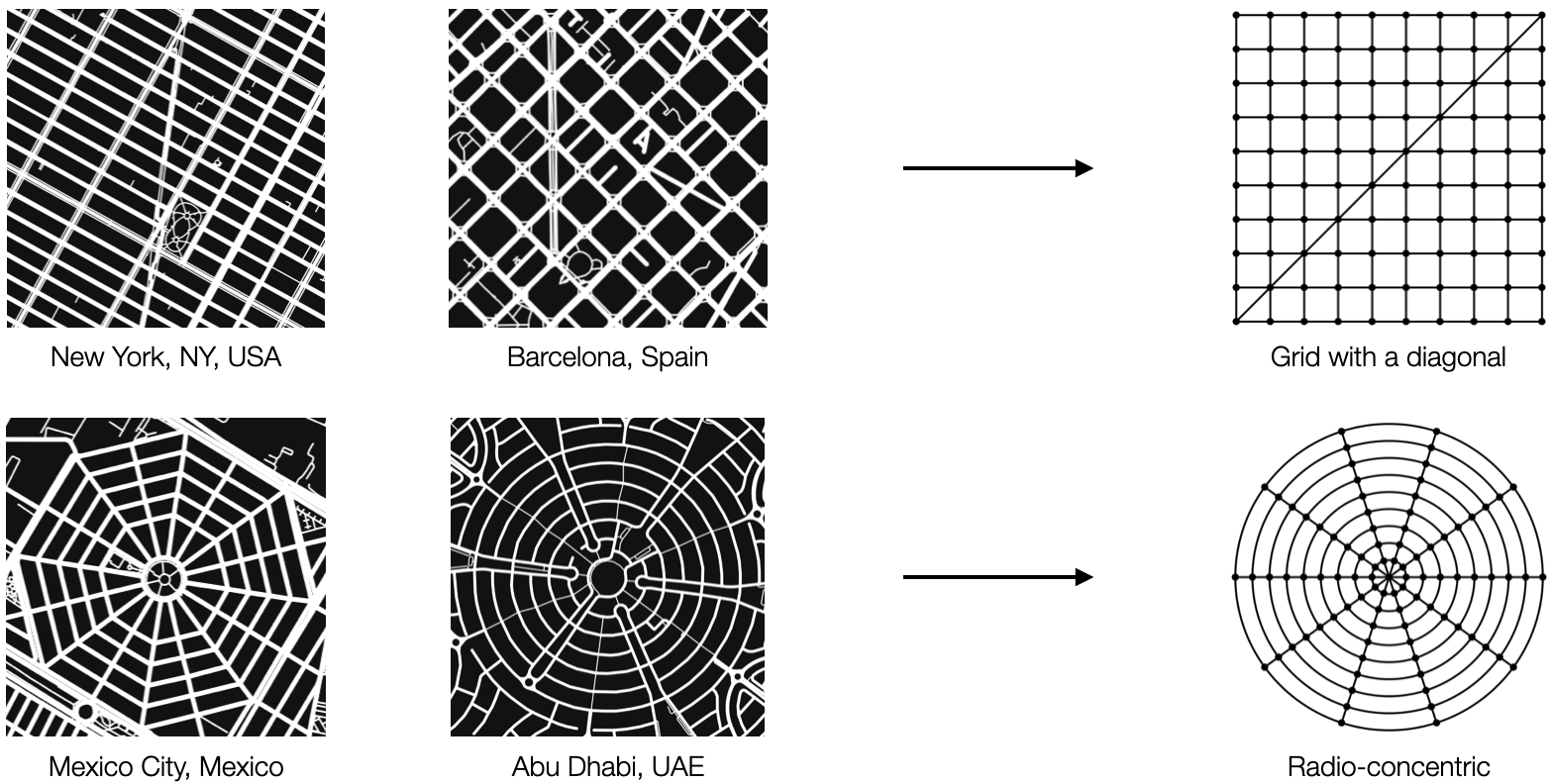}
    \caption{Examples of real street networks and their ideal type.}
    \label{justification-stylized-graph}
\end{figure}

Since this is the most typical use case, we use stylized street networks that are representing typical urban patterns, in line with the urban planning literature. With a connectedness requirement in place, the growth of a tree network would be highly constrained; therefore, we are limiting ourselves to urban patterns that are not trees. We explore four different structures (see Fig. \ref{toy-graph-fig}) and use them as ideal proxies for real street networks, see Fig.~\ref{justification-stylized-graph}. We find that most results are qualitatively similar for non-gridlike structures, so for reasons of simplicity in the main part of the paper we focus on the regular grid.

For access to the repository producing the stylized networks used here, see \cite{github_utg}.

\subsection{Square grid}

The grid, also called rectilinear, gridiron, or Hippodamian plan \citep{lariceUrbanDesignReader2012}, is one of the most ubiquitous shapes found in street networks. From the Centuriation to the contemporary layout of most US cities, it allows for a regular, highly-connected, simple plan. After human mobility systems became increasingly car-centered, the square grid was also celebrated by modernist urban planning as it allows for fast, uninterrupted flow of motorized traffic. As Le Corbusier puts it: ``The circulation of traffic demands the straight line; it is the proper thing for the heart of a city. The curve is ruinous, difficult and dangerous; it is a paralyzing thing. […] The winding road is the Pack-Donkey’s Way, the straight road is man’s way.'' \citep{corbusierCityTomorrowIts1929}.

We motivate the grid size used in the main part of this study by yielding a balanced number of 180 links, allowing to observe growth behavior without the need for long computational times. It also allows to create larger blocks within the grid of 9 blocks of $3\times3$ faces, as in Barcelona's Superblocks. Having an even number of nodes further avoids link formation along the lines of symmetry.

\subsection{Grid with a diagonal}

A variant of the grid is to add a diagonal street along the grid, to allow for better connectivity and straightness through a major street crossing through the grid. Famous examples are the Avinguda Diagonal in Barcelona's neighborhood l'Eixemple, or Broadway in New York City's neighborhood Manhattan.

\subsection{Bridges}

Many cities grew around rivers because of the benefits of having direct access to water: Cairo, Paris, Hangzhou, New York, Budapest, etc. Bridges plays an important role as being the sparse links between parts of the cities. They have been studied since what is usually referred as the first example of applied network science in the ``Seven bridges of Königsberg'' problem of Euler \citep{hopkinsTruthKonigsberg2004}. Bridges are important because they represent a small number of links that connect entire parts of the city, constituting a bottleneck for traffic and being critical for the network's resilience.

\subsection{Radio-concentric}

Another typical urban pattern is the radio-concentric pattern, in which streets emerge radially outwards from a central point of interest (historically a stronghold, a central plaza, or a church). Those radial streets are then connected to one another by multiple concentric streets. Archetypal examples can be found in French ``circulade'' \citep{watteauxPlanRadioquadrilleTerroirs2003} like Bram, or any city physically embodying the ``concentric zone model'' \citep{park1925city}.

\section{Area Under Curve with exponential discounting}
\label{exp-disc}

\begin{figure}[!ht]
    \centering
    \includegraphics[width=\linewidth]{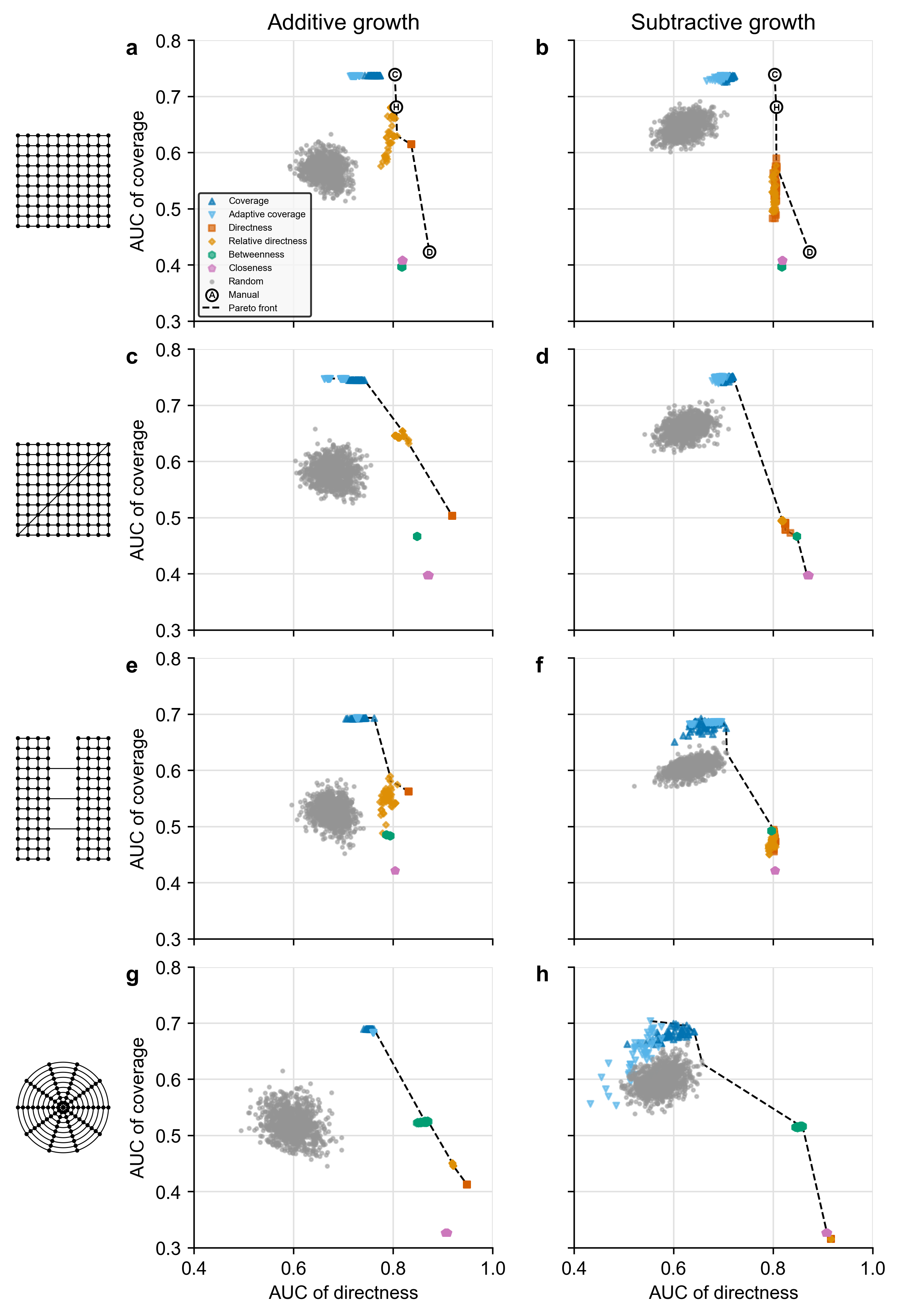}
    \caption{Comparison of the AUC of coverage and directness with exponential discounting for \textbf{a} all additive growth strategies on a square grid, \textbf{b} all subtractive growth strategies on a square grid, \textbf{c} all additive growth strategies on a square grid with diagonal, \textbf{d} all subtractive growth strategies on a square grid with diagonal, \textbf{e} all additive growth strategies on a square grid with bridges, \textbf{f} all subtractive growth strategies on a square grid with bridges, \textbf{g} all additive growth strategies on a radio-concentric network, \textbf{h} all subtractive growth strategies on a radio-concentric network.}
    \label{fig:exp-disc}
\end{figure}

Benefits of transport infrastructure is cumulative: The earlier it is built, the larger are its total benefits for a fixed time horizon. Therefore, we check results for a growth variant where the early stages are prioritized, using exponential discounting. We define the exponentially discounted performance metric $y_g$ as:

\begin{equation*}
    y_g(x,y)=y e^{-x \ln(g)},
\end{equation*}

\noindent where $y$ is the performance metric, $x$ is the normalized built length and $g$ is the discounting gap, where $y_{g}(0,y)/y_{g}(1,y) = g$, meaning that the last link is valued $g$ times less than the first link. We report results in Fig.~\ref{fig:exp-disc} for $g=10$. As focused growth at an early stage is rewarded, we observe a larger gap between random and greedy strategies, exacerbating previous observations. Qualitatively, results remain similar to no discounting, compare with Fig.~\ref{AUC-sm}.

\end{appendices}

\clearpage


\end{document}